# NHANES-GCP: Leveraging the Google Cloud Platform and BigQuery ML for reproducible machine learning with data from the National Health and Nutrition Examination Survey

B. Ross Katz[1], Abdul Khan[1], James York-Winegar[1], and Alexander J. Titus[2, 3, 4]

[1]CorrDyn, Nashville, 37219, USA
[2]In Vivo Group, Los Angeles, 90292, USA
[3]Information Sciences Institute, University of Southern California, Los Angeles, 90292, USA
[4]Iovine and Young Academy, University of Southern California, Los Angeles, 90089, USA

Corresponding author: Alexander J. Titus (publications@theinvivogroup.com)

# Abstract

## Summary

NHANES, the National Health and Nutrition Examination Survey, is a program of studies led by the Centers for Disease Control and Prevention (CDC) designed to assess the health and nutritional status of adults and children in the United States (U.S.). NHANES data is frequently used by biostatisticians and clinical scientists to study health trends across the U.S., but every analysis requires extensive data management and cleaning before use and this repetitive data engineering collectively costs valuable research time and decreases the reproducibility of analyses. Here, we introduce NHANES-GCP, a Cloud Development Kit for Terraform (CDKTF) Infrastructure-as-Code (IaC) and Data Build Tool (dbt) resources built on the Google Cloud Platform (GCP) that automates the data engineering and management aspects of working with NHANES data. With current GCP pricing, NHANES-GCP costs less than $2 to run and less than $15/yr of ongoing costs for hosting the NHANES data, all while providing researchers with clean data tables that can readily be integrated for large-scale analyses. We provide examples of leveraging BigQuery ML to carry out the process of selecting data, integrating data, training machine learning and statistical models, and generating results all from a single SQL-like query. NHANES-GCP is designed to enhance the reproducibility of analyses and create a well-engineered NHANES data resource for statistics, machine learning, and fine-tuning Large Language Models (LLMs).

## Availability and implementation

NHANES-GCP is available at https://github.com/In-Vivo-Group/NHANES-GCP

## Introduction

The National Health and Nutrition Examination Survey (NHANES) has been instrumental in shaping public health policies and research in the United States. It provides a comprehensive dataset reflecting the health and nutritional status of the U.S. population, covering a wide range of demographic groups. Despite its extensive utility, the effective use of NHANES data often requires significant data management and preprocessing (1), which can be both time-consuming and a barrier to reproducibility (2). These challenges necessitate innovative solutions to streamline the research process and enhance data usability.

In response to this need, we introduce NHANES-GCP, a novel infrastructure developed on the Google Cloud Platform (GCP). These Cloud Development Kit for Terraform (CDKTF) and Data Build Tool (dbt) resources automate the data engineering process for NHANES data, addressing the critical need for efficiency and reproducibility in research (3). The operational cost-effectiveness of NHANES-GCP, with a setup cost less than the cost of a cup of coffee (<$2) and minimal ongoing fees (<$15/yr), makes it a viable option for researchers and institutions of varying scales.

A pivotal feature of NHANES-GCP is its integration with BigQuery ML, which simplifies data analysis through an SQL-like interface, allowing for seamless model training and statistical analysis (4). This integration signifies a progressive step towards the amalgamation of traditional statistical approaches and modern machine learning techniques, enhancing the scope and depth of health data analysis.

Crucially, NHANES-GCP addresses the challenge of reproducibility in scientific research, a subject of growing concern in the academic community (5). By standardizing data management and providing clean, structured data, NHANES-GCP ensures that research outcomes can be reliably replicated and validated, contributing significantly to the analysis method's integrity.

Additionally, the structured and quality-assured data from NHANES-GCP offers a valuable resource for the development and training of Large Language Models (LLMs) (6). Such models require high-quality, diverse datasets for effective training (7), and NHANES-GCP facilitates this, opening new avenues in AI and machine learning research, particularly in health and nutrition.

NHANES-GCP represents a modern approach to utilizing NHANES data for health and nutrition research. By harnessing the capabilities of GCP and BigQuery ML, it offers a cost-effective, efficient, and reproducible methodology for data analysis. The potential of NHANES-GCP to streamline research processes and contribute to the advancement of knowledge in the field is substantial, marking a step forward in the utilization of NHANES data for scientific inquiry.

# Implementation

## System Architecture and Prerequisites

NHANES-GCP is structured on a robust and scalable architecture, optimized for handling large datasets typical of NHANES. The system is designed to operate within the GCP ecosystem, utilizing services such as BigQuery for data storage and analysis, and Compute Engine for scalable computing resources.

To set up NHANES-GCP, users must first ensure they have access to GCP with the necessary permissions. The **cdktf.json** file outlines the required modules and configuration for CDKTF, which is instrumental in creating cloud infrastructure by defining and provisioning cloud infrastructure resources in a declarative way using familiar programming constructs.

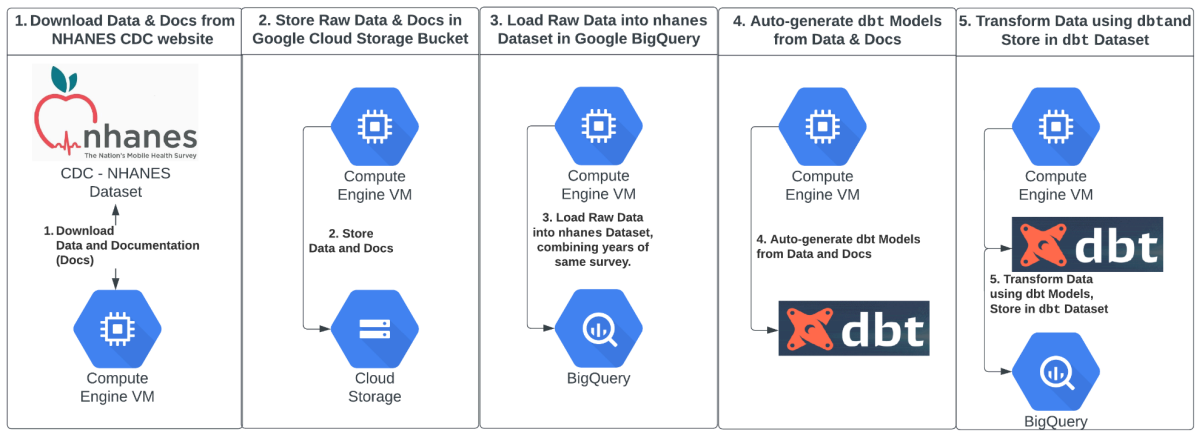

*Figure 1*: The solution architecture diagram for NHANES-GCP. The solution is built using CDKTF to create Infrastructure-as-Code (IaC) to manage GCP resources, allowing the end-user to run a set of simple commands to spin up, manage, and turn off GCP resources as well as access, download, and manage the NHANES data from the CDC website.

## Deployment Process

The deployment of NHANES-GCP is facilitated through a series of scripts and configuration files. The **main.ts** file serves as the entry point, orchestrating the deployment process. This TypeScript file leverages the CDKTF to define and deploy the necessary infrastructure components.

The **startup-script.sh** is a crucial shell script used during the deployment process. It is responsible for initializing and configuring the Compute Engine instances, ensuring they are correctly set up to handle the NHANES data processing tasks. This runs in a virtual machine (VM) as a startup script instead of a Docker container due to chaining multiple processes together and a long run time for extracting data from NHANES.

## Data Processing and Analysis

Once deployed, NHANES-GCP spins up a VM in the GCP environment and runs a series of Python scripts to extract, load, and transform the data. All files listed can be located in the `ELT` directory of the GitHub repository.

First, metadata for each year of each survey is scraped and stored in the `nhanes.nhanes_file_metadata` table in BigQuery (scrape_nhanes_metadata.py). Next, we iterate through the datasets and documentation for each year, downloading each to Google Cloud Storage (GCS) and converting data files from XPT (SAS) to parquet format (scrape_nhanes_data_files.py). Third, we utilize DuckDB to query parquet files in GCS from each survey and combine them together into a single table per survey type, stored in the `nhanes` dataset in the user's BigQuery project.

dbt is an open-source package that provides an increasingly common method of documenting and version-controlling transformations of data that are executed via SQL in data warehouse environments, including BigQuery. We utilize a Python script (generate_dbt_models.py) to auto-generate dbt models for each survey from the documentation downloaded from the NHANES website. Each dbt model contains commented URLs indicating which documentation files were used to generate it. Models are stored in the ./dbt/models/ directory of the provided repository. This process enables NHANES-GCP to be responsive to changes in data format or schema or the addition of new surveys or new years from existing surveys that occur in the future. This also allows the NHANES-GCP resource to maintain consistent and reproducible data transformation for future analyses.

Finally, we utilize the dbt command line interface to build the dbt models in the `dbt` dataset in the user's BigQuery project. To demonstrate analysis of the provided data, we provide example BigQuery ML SQL queries that build logistic regression models and produce results. These examples can be found in the ./dbt/analyses directory.

## Integration with BigQuery ML

One of the standout features of NHANES-GCP is its integration with BigQuery ML. This integration allows researchers to leverage machine learning capabilities directly within the BigQuery environment, facilitating complex data analyses without the need for additional machine learning tools or frameworks.

## Package Management and Dependencies

The **package.json** and **package-lock.json** files manage the dependencies and packages required for NHANES-GCP. These files ensure that all necessary JavaScript packages are correctly installed and maintained, contributing to the stability and maintainability of the project.

## BigQuery ML Example Models

The NHANES-GCP project incorporates several BigQuery Machine Learning (BQML) example models, each designed to analyze specific health-related aspects using logistic regression. These models provide insights into the relationships between various health indicators and outcomes and demonstrate the ease of use for leveraging NHANES-GCP for research. These analyses can be found in the /dbt/analyses directory.

1. **Grip Strength to Depression Model**: This model uses logistic regression to explore the relationship between grip strength (as measured by **combined_grip_strength**) and clinically relevant depression (**has_clinically_relevant_depression**). The depression variable is derived from responses to a depression screener questionnaire, with a score of 10 or more indicating clinically relevant depression. This model offers insights into how physical strength correlates with mental health (8).
2. **Physical Activity Limitation and Food Security Model**: This model investigates the impact of household food security on physical activity limitations. It considers various difficulties related to physical activities, like walking or lifting, and correlates them with the food security status of the household. The logistic regression model aims to understand how nutritional factors affect physical capabilities (9).
3. **Instrumental Activity Limitation and Food Security Model**: Similar to the physical activity model, this model examines the relationship between food security and difficulties in instrumental activities, such as preparing meals or managing money. The model assesses how food security influences the ability to perform these complex daily tasks (9).
4. **Basic Activity Limitation and Food Security Model**: This model explores the connection between food security and limitations in basic daily activities, such as getting in and out of bed or dressing oneself. It aims to identify how food security status impacts the ability to perform basic functional tasks (9).

Each model is implemented in BQML with options like **LOGISTIC_REG** for logistic regression, **input_label_cols** to specify the dependent variable, and **CATEGORY_ENCODING_METHOD = 'DUMMY_ENCODING'** for handling categorical variables. These models leverage BigQuery's powerful ML capabilities to analyze complex relationships within the NHANES dataset efficiently.

Through these models, NHANES-GCP demonstrates a comprehensive approach to understanding the multifaceted nature of health and wellness, illustrating the interplay between physical strength, mental health, and socio-economic factors like food security.

## Conclusion

The NHANES-GCP project, leveraging GCP and BigQuery ML, marks a modern advancement in the field of health data analysis. By automating the data engineering process and integrating advanced machine learning capabilities, NHANES-GCP not only enhances the efficiency and reproducibility of research but also opens new avenues for in-depth health and nutrition studies.

The use of CDKTF has streamlined the deployment and management of cloud resources, making it more accessible and cost-effective for researchers, costing < $2 to deploy and < $15 annually to maintain the data and compute resources. This approach addresses the long-standing challenges of data management in NHANES, reducing the time and resources required for data preparation.

The integration of BigQuery ML stands out as a key feature, allowing researchers to perform analyses, including statistical regression and machine learning models, directly within the cloud environment. This capability is exemplified in the project's demonstration models analyzing relationships between grip strength and depression, as well as food security and various activity limitations. These models have provided valuable insights, demonstrating the interplay between physical, mental, and socio-economic health factors.

Looking forward, NHANES-GCP sets a new standard for health data analysis. Its framework can be adapted for other large-scale health datasets, potentially transforming research methodologies across various domains. The project also offers a robust platform for leveraging the NHANES dataset for training and fine-tuning Large Language Models (LLMs), contributing to advancements in artificial intelligence and machine learning in health research.

NHANES-GCP embodies a modern approach to utilizing health data, combining technical innovation with practical application. It offers a scalable, reproducible, and efficient solution for researchers, paving the way for future health and nutrition research. As the field continues to evolve, NHANES-GCP is a resource for integrating cloud computing and machine learning technologies in scientific inquiry.

## Acknowledgments

None declared.

## Supplementary Data

None declared.

## Conflicts of Interest

None declared.

## Funding

None declared.

## Data Availability

The NHANES-GCP code is freely available at https://github.com/In-Vivo-Group/NHANES-GCP and the NHANES data can be found at https://www.cdc.gov/nchs/nhanes/index.htm.